# Comparison of Compensation Mechanism Between an NMR Gyroscope and an SERF Gyroscope

Haifeng Dong and Yang Gao

*Abstract*—Through analysis of the compensation mechanism of nuclear-magnetic-resonance (NMR) gyroscope and spin-exchange-relaxation-free (SERF) gyroscope, we demonstrate that there is a common model for these two kinds of an atomic rotation rate sensor. The output signals of NMR and SERF gyroscopes are compared directly, which provides a guidance for the scheme choosing and optimization of atomic gyroscope. The input–output relation of both gyroscopes is given, which can be used to analyze the contributions of different error sources.

*Index Terms*— Atomic gyroscope, nuclear magnetic resonance, spin exchange relaxation free.

## I. INTRODUCTION

GYROSCOPES are used in both practical applications such as navigation and positioning and fundamental physics [1], [2]. Nuclear magnetic resonance (NMR) and spin exchange relaxation free (SERF) gyroscope are two kinds of actively-developed nuclear-spin gyroscope schemes [3].

In a typical NMR gyroscope configuration, two nuclear spin species are used to suppress magnetic field noise. One of them is typically used to stabilize the magnetic field while the rotation rate is measured through the other [4]. In recent experiments, alkali vapor is used to pump the nuclear spin through spin-exchange and to probe the Larmor precession of the nuclear spin polarization [5]–[7]. The Fermi-contact interaction between alkali atoms and nuclear spin species enhances the signal, while it also leads to the bias drift. The methods to avoid this drift such as pulsed pumping and detection [8], synchronous spin-exchange optical pumping [9] and averaging Rb polarization using $\pi$ pulse to eliminates the back-polarization [10] are beyond the topic of this paper. Here we focus on the compensation mechanism of NMR as well as SERF gyroscope and provide a direct comparison of these atomic gyroscopes.

For the SERF gyroscope, only one nuclear spin species is used. It is also pumped by spin-exchange with alkali atoms. However, besides the pumping effect, the electron spin of

Manuscript received April 5, 2017; accepted May 8, 2017. Date of publication May 11, 2017; date of current version June 12, 2017. This work was supported in part by the National Natural Science Foundation of China under Grant 51675034 and Grant 61273067 and in part by the Science Foundation of Beijing Municipality under Grant 7172123. The associate editor coordinating the review of this paper and approving it for publication was Dr. Minghong Yang. *(Corresponding author: Haifeng Dong.)*

The authors are with the School of Instrumentation Science and Opto-Electronics Engineering, Beihang University, Beijing 100191, China, also with the Science and Technology on Inertial Laboratory, Beijing 100191, China, and also with the Fundamental Science on Novel Inertial Instrument and Navigation System Technology Laboratory, Beijing 100191, China (e-mail: hfdong@buaa.edu.cn; gaoyang17@buaa.edu.cn).

Digital Object Identifier 10.1109/JSEN.2017.2703601

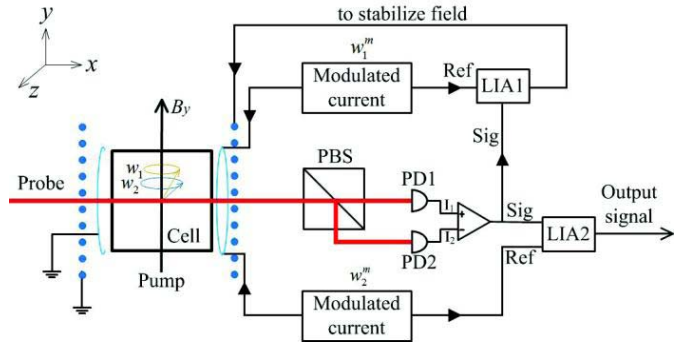

Fig. 1. The measurement scheme for a NMR gyroscope. PBS: polarizing beam splitter, PD: photodetector, LIA: lock-in amplifier

alkali atom is also used to sense the rotation directly instead of as a precession prober. Furthermore, unlike in the NMR scheme where the output is the nuclear spin polarization precession rate, the output of this scheme is the steady-state of the alkali polarization projection in the probe direction, which limits its response bandwidth to about $1/T_{2a}$, where $T_{2a}$ is mainly decided by the electron spin destruction relaxation of alkali atom in the SERF regime [11].

Under the superficial difference, two gyroscopes have commonality in essence. In section II, we analyze the compensation mechanism of NMR gyroscope. In section III, we analyze that of SERF gyroscope. In section IV, we compare the compensation equation, enhancement factor and the output signal of two gyroscopes directly. The conclusion is give in section V.

## II. NMR GYROSCOPE ANALYSIS

Fig. 1 shows the basic measurement scheme for a NMR gyroscope. Ignoring the alkali magnetization and the quadruple effect, the equivalent total fields experienced by the first nuclear spin is as below [6], [12],

$$B_1 = \frac{w_1}{\gamma_{n1}} = B_y + \frac{\Omega_y}{\gamma_{n1}} \quad (1)$$

where $B_1$ is the equivalent magnetic field experienced by the first nuclear spin, $B_y$ is magnetic field added using coils along $y$-axis, $w_1$ is the Larmor frequencies of the first nuclear spin, $\gamma_{n1}$ is the gyromagnetic ratios of the first nuclear spin, and $\Omega_y$ is the frame rotation rate around $y$-axis. The Larmor precession $w_1$ is measured and servoed to the field coils to stabilize $B_1$. Once $B_1$ is stabilized, equation (1) becomes a compensation equation, which can be rewritten as $B_y = B_1 - \Omega_y/\gamma_{n1}$.

Signal before Lock-in amplifier two ( LIA2 in Fig. 1) is

$$S_b = 2I_0\theta = I_0 n l c r_e f D(v) P_x \quad (2)$$

where $I_0$ is the amplitude of probe beam, $n$ is the density of alkali atom, $l$ is the optical path length, $c$ is the light speed, $r_e$ is the classical radius of the electron, $f$ is the oscillator strength, $D(v) = \frac{v-v_0}{(v-v_0)^2+(\Delta v/2)^2}$ is the absorption Lorentz function, $P_x$ is the steady-state of the electron spin of alkali atom polarization projection in the probe direction as the Larmor precession of nuclear magnetization is adiabatical. As the reference frequency of LIA2 is $w_2$, we neglect the harmonic of $w_1$ in the expression of $P_x$.

$$\begin{aligned}P_x &= P_0\frac{\frac{1}{T_{2a}\gamma_a}B_{xz}\sin(w_2t+\psi)+B_{xz}\cos(w_2t+\psi)B_y}{[B_{xz}\sin(w_2t+\psi)]^2+[B_{xz}\cos(w_2t+\psi)]^2+B_y^2+\left(\frac{1}{T_{2a}\gamma_a}\right)^2}\\ &\approx P_0\frac{B_{xz}}{B_y}\cos(w_2t+\psi)\end{aligned} \quad (3)$$

where $T_{2a}$ is the transverse relaxation time of the electron spin of alkali atom polarization, $\gamma_a$ is the gyromagnetic ratio of alkali atom, $B_{xz}$ is the projection of the second nuclear spin magnetization sensed by alkali atoms in $xz$ plane, $w_2$ is the Larmor precession frequency of the second nuclear spin magnetization, $w_2t + \psi$ is the angle between $B_{xz}$ and $x$-axis, $\psi$ is the initial phase of the nuclear Larmor precession, i.e. the phase difference between modulated current and the Larmor precession of the nuclear magnetization. The approximation in equation (3) is based on the assumption that $B_y$ is much larger than $B_{xz}$ and $1/T_{2a}\gamma_a$.

After Lock-in amplifier, the quadrature output of LIA2 is as below,

$$\begin{aligned}S_{NMR} &= I_0 n l c r_e f D(v) P_0 \frac{B_{xz}}{B_y}\sin(\psi)\\ &\approx I_0 n l c r_e f D(v) P_0 \frac{B_{xz}}{B_y}\psi\\ &= I_0 n l c r_e f D(v) P_0 \frac{B_{xz}}{B_y}T_{2n2}[w_2^m - (B_y\gamma_{n2}+\Omega_y)]\end{aligned} \quad (4)$$

where $T_{2n2}$ is the transverse relaxation time of the second nuclear spin.

Insert the compensation equation (1) into the equation (4), set $w_2^m = B_1\gamma_{n2}$, and use the optimization of $B_{xy} = \frac{B_{n2}}{2}\sqrt{\frac{T_{2n2}}{T_{1n2}}}$ at zero detune, we obtain

$$\begin{aligned}S_{NMR} &= I_0 n l c r_e f D(v) P_0 \frac{B_{xz}}{B_y}\left[T_{2n2}\left(\frac{\gamma_{n2}}{\gamma_{n1}}-1\right)\Omega_y\right]\\ &\approx I_0 n l c r_e f D(v) P_0 \frac{B_{n2}}{B_y}\frac{1}{2}\sqrt{\frac{T_{2n2}}{T_{1n2}}}T_{2n2}\left(\frac{\gamma_{n2}}{\gamma_{n1}}-1\right)\Omega_y\\ &= \frac{1}{2}\frac{B_{n2}}{B_y}\sqrt{\frac{T_{2n2}}{T_{1n2}}}\cdot H\cdot T_{2n2}\left(\frac{\gamma_{n2}}{\gamma_{n1}}-1\right)\Omega_y\end{aligned} \quad (5)$$

where $B_{n2}$ and $T_{1n2}$ are the magnetization and the longitudinal relaxation time of the second nuclear spin, respectively, $H = I_0 n l c r_e f D(v)$.

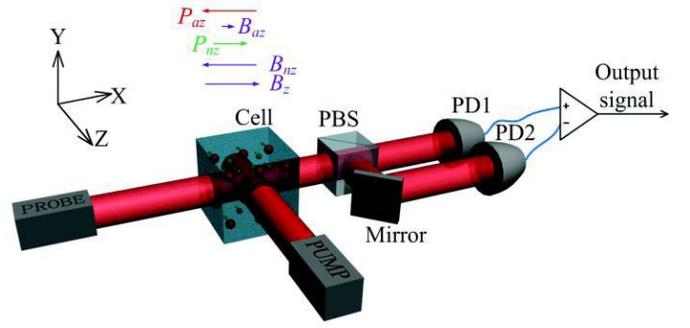

Fig. 2. The measurement scheme for a SERF gyroscope. The polarization and magnetization of alkali and noble nuclear, as well as the compensation field in $z$ direction are shown inset. PBS: polarizing beam splitter, PD: photodetector.

The approximations in equation (4) and (5) are under the assumption that the modulated frequency is close to the nuclear polarization Larmor precession frequency so that $\psi$ is close to zero and $B_{xz} \approx B_x$.

From equation (5), we can see that there is no bias drift related to $B_y$ in the output signal, while variations of $B_y$ do lead to the scale factor instability. And $\Omega_y$ is enhanced by a factor of $\frac{\gamma_{n2}}{\gamma_{n1}} - 1$.

### III. SERF GYROSCOPE ANALYSIS

Fig. 2 is the basic measurement scheme for a SERF gyroscope. $B_x$ and $B_y$ are compensated to zero using magnetic shield and coils (not shown in the figure). In this way the polarization of alkali can be regarded as in $z$ direction. Thus the spin-exchange pumping direction of the noble gas is also in $z$ direction and the steady-state nuclear polarization projection in $y$-axis is,

$$\begin{aligned}P_{ny} &= P_{n0}\frac{\frac{1}{T_{2n}\gamma_n}B_x + (B_y+\Omega_y/\gamma_n)(B_z-B_{az})}{B_x^2+(B_y+\Omega_y/\gamma_n)^2+(B_z-B_{az})^2+\left(\frac{1}{T_{2n}\gamma_n}\right)^2}\\ &\approx P_{nz}\frac{B_y+\Omega_y/\gamma_n}{B_z-B_{az}}\end{aligned} \quad (6)$$

where $P_{n0}$ is the nuclear polarization, $P_{nz}$ is the nuclear polarization projection in $z$-axis, $T_{2n}$ and $\gamma_n$ are the transverse relaxation time and the gyromagnetic ratio of the nuclear, respectively, $B_z$ is added using coils to compensate nuclear magnetization $B_{nz}$ so that the alkali works under SERF regime. Setting $B_z = B_{az} + B_{nz}$ and considering that $B_{nz} = kP_{nz}$, where $k$ is the ratio between nuclear polarization and magnetization, we can obtain the compensation equation of SERF gyroscope from equation (6),

$$B_y = kP_{ny} - \Omega_y/\gamma_n \quad (7)$$

The differential polarimetry output signal in Fig. 2 is

$$S_{SERF} = 2I_0\theta = I_0 n l c r_e f D(v) P_x \quad (8)$$

where under the SERF regime, $P_x$ can be obtained from the steady-state solution of the Bloch equation,

$$P_x \approx -P_0 T_{2a}\gamma_a B_{ay} \quad (9)$$

TABLE I
COMPARISON BETWEEN NMR AND SERF GYROSCOPE

| | Compensation equation | Output Signal | Enhancement factor |
|---|---|---|---|
| NMR gyroscope | $B_y = B_1 - \Omega_y / \gamma_{n1}$ | $\frac{1}{2}\frac{B_{n2}}{B_y}\sqrt{\frac{T_{2n2}}{T_{1n2}}} \cdot H \cdot T_{2n2}\left(\frac{\gamma_{n2}}{\gamma_{n1}}-1\right)\Omega_y$ | $\frac{\gamma_{n2}}{\gamma_{n1}}-1$ |
| SERF gyroscope | $B_y = kP_{ny} - \Omega_y / \gamma_n$ | $H \cdot T_{2a}\left(\frac{\gamma_a}{\gamma_n}-1\right)\Omega_y$ | $\frac{\gamma_a}{\gamma_n}-1$ |

where $T_{2a}$ and $\gamma_a$ are the transverse relaxation time and the gyromagnetic ratio of the alkali atoms, respectively, $B_{ay} = B_y + \Omega_y/\gamma_a - kP_{ny}$ is the effective field sensed by the alkali atoms.

Insert equation (9) into equation (8) and replace $B_{ay}$ with $B_y + \Omega_y/\gamma_a - kP_{ny}$, we obtain,

$$S_{SERF} = I_0 n l c r_e f D(v) P_0 T_{2a}[\gamma_a k P_{ny} - (\gamma_a B_y + \Omega_y)] \quad (10)$$

Insert the compensation equation (7) into (10) equation, we obtain,

$$S_{SERF} = I_0 n l c r_e f D(v) P_0 T_{2a}(\frac{\gamma_a}{\gamma_n} - 1)\Omega_y$$
$$= H \cdot T_{2a}(\frac{\gamma_a}{\gamma_n} - 1)\Omega_y \quad (11)$$

## IV. COMPARISON AND DISCUSSION

Table I compares NMR gyroscope and SERF gyroscope. Both kinds of gyroscope has a compensation equation with the same form. $B_y$ is the bridge between two kinds of species. For the compensation specie there must be an equation between $B_y$ and the equivalent field of rotation, which becomes a compensation equation once the extra term can be stabilized ($B_1$ in NMR) or eliminated ($kP_{ny}$ in SERF). The output signals of both gyroscopes are also similar. There are no bias drifts corresponding to the magnetic field variations and both signals are enhanced by a factor decided by the ratio of gyromagnetic ratios of the coupled species.

The differences are that: Firstly, because the $\gamma_a$ is usually three orders larger than $\gamma_n$, the enhancement factor of a SERF gyroscope is usually bigger than that of a NMR gyroscope. On the contrary, the longer transverse relaxation time of the nuclear spin polarization can increase the NMR signal to offset this shortage. Secondly, there is an extra terms in NMR gyroscope output signal, i.e. $\frac{1}{2}\frac{B_{n2}}{B_y}\sqrt{\frac{T_{2n2}}{T_{1n2}}}$, from which we can see that the signal is proportional to the ratio of $B_{n2}$ and $B_y$ and that although $B_y$ in NMR gyroscope is eliminated from the bias drift after compensation, it still affects the scale factor in an inversely-proportional form.

To compare NMR and SERF gyroscope further, we use the typical parameters of Cs-Xe NMR and SERF gyroscopes (listed in Table II) and calculate the scale factor of both kinds of gyroscope using equation (5) and equation (11). In equation (11) $B_{n2}$ is calculated by $B_{n2} = \frac{8\pi \kappa_0 \mu_0}{3}\mu n_{Xe} P$, where $\kappa_0$ is the enhancement factor for CsXe, $\mu_0$ is vacuum permeability, $\mu$ is atom magnetic moment, $n_{Xe}$ is atom density of 129Xe and $P$ is the polarization of 129Xe. For NMR gyroscope, 131Xe is used as the first nuclear spin species and 129Xe is used as the second nuclear spin species. For SERF gyroscope, 129Xe is used to generate the nuclear polarization. $H = I_0 n l c r_e f D(v)$ is reprinted here for reference where $I_0$ is calculated using probe optical power of 1mW and responsivity of 0.6mA/mW corresponding to Newfocus Model2307 photodetector. $D(v)$ is calculated at the half linewidth detune and the linewidth $\Delta v$ is calculated according to the pressure broadening. The results show that the scale factor of NMR gyroscope is about $4.05 \cdot 10^{-3}$ mA/°/h while that of SERF gyroscope is about $6.24 \cdot 10^{-3}$ mA/°/h. Considering the technical 1/f noise from laser and photodiode [13], [14], the corresponding bias stabilities of NMR and SERF gyroscope are estimated to be 0.003°/h and 0.002°/h, respectively.

TABLE II
TYPICAL PARAMETERS OF Cs-Xe ATOMIC ENSEMBLE

| Parameter | Variable | unit |
|---|---|---|
| enhancement factor | $\kappa_0$ | 880 [15] |
| vacuum permeability | $\mu_0$ | $4\pi \cdot 10^{-7}$ Wb/A·m |
| 129Xe magnetic moment | $\mu$ | $0.84 \cdot 10^{-26}$ J/T [16] |
| 129Xe atom density | $n_{Xe}$ | $9.38 \cdot 10^{22}$ m$^{-3}$ [8] |
| 129Xe polarization | $P$ | 2% [17, 18] |
| coil field along y-axis | $B_y$ | 2000nT |
| longitudinal relaxation time of the second nuclear spin | $T_{2n2}$ | 16.4s [19] |
| transverse relaxation time of the second nuclear spin | $T_{1n2}$ | 60s [18] |
| amplitude of the probe beam after photodetector | $I_0$ | 0.6mA |
| Cs atom density at 120°C | $n$ | $4.91 \cdot 10^{19}$ |
| optical path length | $l$ | 1cm |
| speed of light | $c$ | $3 \cdot 10^8$ m/s |
| classical radius of the electron | $r_e$ | $2.82 \cdot 10^{-15}$ m [20] |
| oscillator strength | $f$ | 0.347 |
| absorption linewidth | $\Delta v$ | 66.23MHz |
| 129Xe gyromagnetic ratio | $\gamma_{n2}$ | -11.86MHz/T [3] |
| 131Xe gyromagnetic ratio | $\gamma_{n1}$ | 3.52MHz/T [3] |
| transverse relaxatiaon time of Cs | $T_{2a}$ | 3.33ms [20] |
| Cs gyromagnetic ratio | $\gamma_a$ | 3500MHz/T [3] |
| 129Xe gyromagnetic ratio | $\gamma_n$ | -11.86MHz/T [3] |

## V. CONCLUSION

We analyze the commonality between two actively-developed nuclear-spin gyroscope schemes. The results show that both gyroscopes share similar compensation mechanism, which may be used in the further improvement of atomic gyroscope, such as the elimination of the alkali magnetization. The quantitative and direct comparison of the output signal provides a guidance for the scheme choosing and optimization of atomic gyroscope. And the analysis also gives a concise view how the self-compensation in SERF gyroscope works and how the main field $B_y$ affects the scale factor in the NMR gyroscope even after compensation. The input-output relation can be used to analyze the contributions of different error sources.

## ACKNOWLEDGMENT

The authors would like to thank Prof. Luo Jun, Dr. Zongmin Ma, Dr. Jun Tang and Dr. Hao Guo for beneficial discussion, which play an important role in the formation of the comparison idea. The authors also thank Dr. Morgan Hedges for his helpful modification and suggestion of the manuscript.

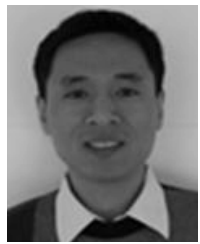

**Haifeng Dong** was born in Shanxi, China, in 1973. He received the Ph.D. degree in microelectronics and solid electronics from Peking University, China, in 2005. He held a post-doctoral position with the Massachusetts Institute of Technology, USA. He is currently a Visiting Professor with the Physics Department, Princeton University. He is currently an Associate Professor at Beihang University. He has presided over three National Natural Science Foundations of China and published papers in *Applied Physics Letters*, *Sensors and Actuators*, the IEEE SENSORS JOURNAL, *Microsystem Technology*, *Optical Communications*, and the *European Physical Journals* as the first and/or corresponding author. His research interests include atomic sensors based on atom spin, micronano fabrication process, and MEMS devices.

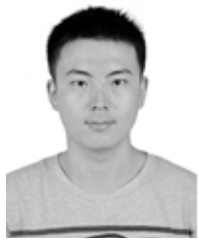

**Yang Gao** was born in Shanxi, China, in 1992. He received the bachelor's degree in measurement and control technology and instrument from the Taiyuan University of Technology, China, in 2015. He is currently a Post-Graduate Student in precision instrument and machinery with Beihang University. His research interests include high sensitive atomic spin polarization sensing and its application.